\documentclass[aps,prl,onecolumn,preprint,floatfix,showpacs]{revtex4}

\usepackage{psfig}
\usepackage{color}

\newcommand{\ea}{\textit{et al.}}

\begin{document}

\title{Constraints on Topological Defects Energy Density from 
First-Year WMAP Results} 

\author{Aur\'elien A. Fraisse}

\email{fraisse@astro.princeton.edu}

\affiliation{Department of Astrophysical Sciences, Princeton
  University, Princeton, NJ 08544}

\date{April 12, 2005}
%------------------------------------------------------------------------------

\begin{abstract}

We compare the predictions of hybrid inflationary models that produce
both adiabatic fluctuations and topological defects to first year WMAP
results. We use a Markov Chain Monte Carlo method to constrain the
contribution of cosmic strings and textures to the CMB angular power
spectrum. Marginalizing this contribution over the cosmological
parameters of a power law flat $\Lambda$CDM model, we place a 95\%
upper limit of 23\% on the topological defects contribution to density
fluctuations, the maximum likelihood being of order 4\%. This
corresponds to an upper limit on the string scale of $G\mu \leq 3.2
\times 10^{-7}$. We also explore the degeneracies between the defects
contribution and other cosmological parameters.  

\end{abstract}

%------------------------------------------------------------------------------

\maketitle

\begin{center}
\textbf{I. \quad INTRODUCTION}
\end{center}

In the 1980's and 1990's, cosmologists focused on two plausible
scenarios, each of which produces scale invariant density
fluctuations. In the inflationary model, quantum fluctuations are
amplified during inflation and produce adiabatic, gaussian and nearly
scale invariant fluctuations \cite{Linde}. This model is successful in
resolving some issues of the Hot Big Bang Model \cite{Dodelson} and
is in good agreement with observations over a wide range of scales
\cite{WMAPbestfit}.

The alternative defects scenario rests on the realization that since
the Universe has steadily cooled down since the Planck time,
spontaneous breaking of symmetries must have occurred. Each symmetry
breaking may lead to the creation of topological defects such as
monopoles, textures, domain walls, or cosmic strings
\cite{Vilenkin}. A non negligible contribution of monopoles and domain
walls is
ruled out by some basic observations \cite{Srivastava}. However,
textures, which appear when a non-Abelian symmetry is spontaneously
and completely broken, and cosmic strings, due to a $U(1)$ symmetry
breaking phase transition, are both liable. Even if they are not the
dominant source of fluctuations, they may still make detectable
contributions to the CMB and large scale structures formation.

Hybrid inflationary models, \textit{e.g.}, D- and F-term inflationary
models, predict that there should be both quantum fluctuations,
occurring during the inflationary phase, and topological defects,
created at the end of inflation \cite{Kibble}, which also induce
anisotropies in the CMB by gravitational effects \cite{Kolb}.

Some recent papers \cite{Dterm} on these models conclude that cosmic
strings should be responsible for at least 50\%, and at most 85\%, of
the amplitude of fluctuations in the CMB anisotropies power spectrum.
However, other works \cite{Rocher} give a much smaller contribution of
cosmic strings for the same models, and the contribution of
topological defects is highly constrained \cite{Bouchet} (best fit
with a 18\% contribution) from measurements previous to WMAP, in
favor of adiabatic fluctuations.

In this Letter, we study the constraints coming from first year WMAP
results. WMAP has indeed given unprecedented precise results on
fluctuations in the CMB. It is therefore worth studying how models
involving topological defects can fit its data, and what are the
foldings of the constraints hereby established on existing
inflationary models, previous constraints, and cosmological
parameters.

\begin{center}
\textbf{II. \quad MODEL AND METHOD}
\end{center}

\begin{figure}
\centerline{\psfig{file=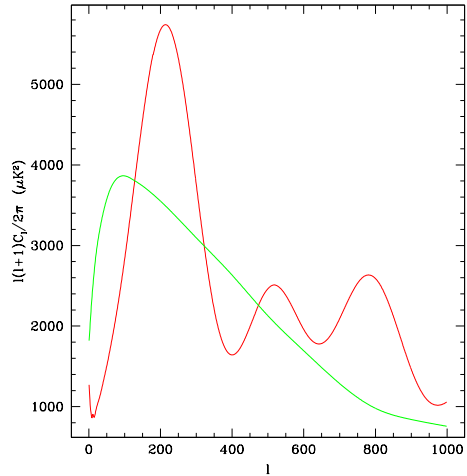,height=2.7in,angle=0}}
\caption{\label{fig:powerspectra} CMB angular power spectrum induced
  by cosmic strings and textures alone (green) and 
  by a simple power law flat $\Lambda$CDM model
  corresponding to WMAP best fit (red). We normalize these spectra so
  that the matter power spectrum normalization for our model is the
  same as WMAP best fit for a flat $\Lambda$CDM model without
  topological defects. It should be noticed that the points of the
  green plot have not been recomputed but directly extracted from
  the study of Seljak, Pen and Turok~\cite{seljak}.}
\end{figure}

As a non negligible contribution of 
monopoles and domain walls is ruled
out, we only
consider cosmic strings (CS) and textures (TX) in this work and  we
assume a model, in which the power spectrum of the CMB anisotropies is
written:
\begin{equation}
\label{eqn:modelinit}
C_{\ell}^{model}=
\left(1-\beta-\gamma\right)\,C_{\ell}^{AD}
+ \beta\,C_{\ell}^{CS}
+ \gamma\,C_{\ell}^{TX}
\end{equation}
\noindent where $\left(\beta,\gamma\right) \in \left[0;1\right]^2$,
$\beta + \gamma \leq 1$ and $AD$ stands for \textit{adiabatic}. The
theoretical power spectra of temperature, and electric type
polarization for strings and textures have already been computed by
Seljak, Pen \& Turok \cite{seljak}. Many tests \cite{seljaktest} have
been performed on the procedure leading to these results, and other
independent works \cite{allen} have given similar conclusions, all of
them being consistent to within about 10\%. These results suggest
that, for $\ell \leq 1000$, the amplitude of fluctuations in the power
spectra of electric type polarization and magnetic type polarization
for cosmic strings and textures, can be safely neglected compared to
the one in the power spectrum of temperature. Moreover, the predicted
power spectra for cosmic strings and textures are so close for these
values of $\ell$ that it is reasonable to consider that they are the
same. Thus, we assume that $C_{\ell}^{CS}$ and $C_{\ell}^{TX}$ are the
one shown on Fig. \ref{fig:powerspectra}, and we can simplify equation
(\ref{eqn:modelinit}) to
\begin{equation}
 \label{eqn:model}
C_{\ell}^{model}=
\left(1-\alpha\right)\,C_{\ell}^{AD}
+ \alpha\,C_{\ell}^{TD}
\end{equation}
\noindent where $C_{\ell}^{TD}$ can be  $C_{\ell}^{CS}$ as well as
$C_{\ell}^{TX}$, and $0\!\leq\!\alpha\nobreak\leq\nobreak 1$ is the
contribution of topological defects to the CMB anisotropies power
spectrum.

To generate the adiabatic spectrum, we use CMBwarp \cite{cmbwarp}, a
fast CMB code that agrees well with the more accurate CMBFast
results. In the parameter range of interest, the largest differences
between these two codes are generally less than 1\%. For the
calculations in this paper, we assume a simple power law flat
$\Lambda$CDM model. We choose to normalize both $C_{\ell}^{AD}$ and
$C_{\ell}^{TD}$ so that, whatever the values of $\alpha$, the matter
power spectrum normalization is the one corresponding to WMAP best fit
as computed by CMBwarp with WMAP best fit cosmological parameters
\cite{WMAPbestfit}.

Thus, our model is chosen as a flat $\Lambda$CDM model to which we add
the contribution of topological defects, so that it can be described
with only 7 parameters: the baryon density ($\omega_b$), the matter
density ($\omega_m$), Hubble constant ($h$), the perturbation
normalization ($A$), the optical depth ($\tau$), the spectral index
($n_s$), and the contribution of topological defects to the power
spectrum ($\alpha$). The idea is to estimate the values of each of
these seven parameters with 1- 2- and 3-$\sigma$ confidence contours
in the 7-D parameter space. 

To investigate the likelihood space, we use a Markov Chain Monte Carlo
(MCMC) method \cite{christensen}, which enables us to evaluate the
likelihood of $\sim 5 \times 10^5$ models in approximatively 7 hours. This
method is based on Bayes' theorem, which states that the probability
to have the set of parameters $s$ given the observed
$\mathcal{C}_\ell$ is
\begin{equation}
\mathcal{P}(s|\mathcal{C}_{\ell})=\frac{\mathcal{P}(\mathcal{C}_{\ell}|s)\,
\mathcal{P}(s)}{\int{\mathcal{P}(\mathcal{C}_{\ell}|s)\,\mathcal{P}(s)\, 
ds}}
\end{equation}
\noindent where $\mathcal{P}(\mathcal{C}_{\ell}|s)$ is the likelihood
of observing $\mathcal{C}_{\ell}$ given the set of parameters $s$, and
$\mathcal{P}(s)$ the prior probability density. This theorem tells us
we can figure out the value of $\mathcal{P}(s|\mathcal{C}_{\ell})$,
what we would like to do, from the likelihood of obtaining the
observed $\mathcal{C}_{\ell}$ from our set of parameters $s$, that we
can compute with the likelihood code provided by the NASA/WMAP Science
Team on LAMBDA, provided we specify priors. Postulating the
equidistribution of ignorance, we assume that each cosmological
parameter has the same probability to have each value between the
following lower and upper bounds:
\begin{center}
$
\begin{array}{cc}
0 \le \omega_b \le 1 & 0 \le \omega_m \le 1 \\
0.5 \le h \le 1.5 & 0.5 \le A \le 2.5 \\
0 \le \tau \le 0.3 & 0 \le n_s \le 2
\end{array}
$
\end{center}
\noindent It should be noticed, as we will see later, that these
priors have no effect on our results, as the MCMC never encounters the
priors' boundaries, at the exception of $\tau$. For this latter, the
prior enables the MCMC not to enter unphysical regions of the
parameter space.

Given these priors and Bayes' theorem, we can then explore the
likelihood surface by computing the likelihood of the model obtained
after each step,  until we find the most likely region for our
cosmological parameters to be. We use the algorithm of \cite{Verde}
for this exploration, and the chain is stopped when we can make sure
it has properly converged (we will study this criterion in details as
a check of our results), which means we have reached and stayed long
enough in the region of the parameters space of highest likelihood, and that 
the parameter space has been widely explored.

\begin{figure}
\centerline{\psfig{file=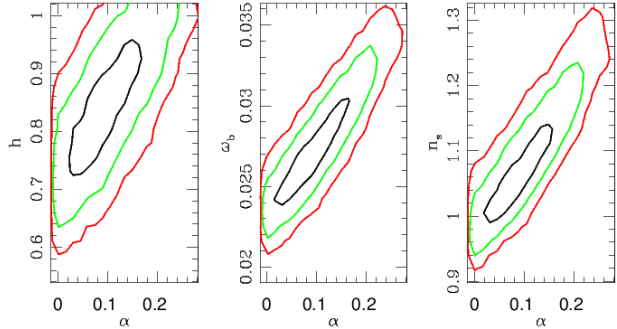,width=3.2in,angle=0}}
\caption{\label{fig:contours} We plot the 1-$\sigma$ (black),
  2-$\sigma$ (green) and 3-$\sigma$ (red) confidence contours for
  first year WMAP data for three parameters: $h$, $\omega_b$ and
  $n_s$. This indicates that scenarios involving a contribution of
  topological defects to the amplitude of fluctuations in the CMB
  greater than 29\% are rejected at 3-$\sigma$, whereas a
  zero-contribution is possible at 2-$\sigma$. The 1-$\sigma$ contour
  implies a contribution between 2\% and 15\%, with a maximum
  likelihood of 4\%. Each parameter on this graph is degenerated with~$\alpha$.}
\end{figure}

In this work, we consider the situations in which the TT and TE
components of the power spectrum are affected by topological
defects. Considering the TT data only, or the TT and TE components
of the power spectrum, does not change the results, among which
is~Fig.~\ref{fig:contours}.

\begin{center}
\textbf{III. \quad RESULTS AND DISCUSSION}
\end{center}

These results show that $\alpha$ is more likely to be found between
0.02 and 0.15 at 1-$\sigma$ with a maximum likelihood around 0.04, but
scenarios involving a zero contribution of topological defects are
authorized at 2-$\sigma$. It is to be noticed that our priors let
$\alpha$ free to take any values. Nevertheless, we see that a
contribution of more than 29\% is completely excluded at
3-$\sigma$. Moreover, we find three main degeneracies between our
parameters: when the value of $\alpha$ increases, so do the values of
$\omega_b$, $n_s$ and $h$. Each of these parameters increases linearly
with~$\alpha$.

This can be easily understand by looking at the respective shapes of
$\mathcal{C}_{\ell}^{TD}$ and $\mathcal{C}_{\ell}^{AD}$
\cite{WMAPbestfit}. Indeed, the spectrum induced by topological
defects gives a very low contribution to the second peak. Therefore, if
we keep the cosmological parameters constant and $\alpha$ increases,
which means that the adiabatic contribution goes down, the height of
the second peak will diminished, so that it will become incompatible
with WMAP results. As our program looks for the most likely set of
cosmological parameters compatible with WMAP, it has to modify the
values of the cosmological parameters, in a way that this discrepancy
is canceled. In other words, the cosmological parameters should be
modified in a way that the height of the second peak increases whereas
the one of the first peak does not change significantly. That is
done by decreasing $\Omega_b$ (which makes the first peak decreases
and the second peak increases) and increasing $h$ (which makes the
first and second peak increases). We can check this behavior of $h$
directly on Fig. \ref{fig:contours}. To verify that the
evolution of $\Omega_b$ is the one predicted, we have to consider the
graph giving $\omega_b$, where we can see that $\omega_b$ increases with
$\alpha$, but less than $h^2$ does, which confirm the decrease of
$\Omega_b$.
 
Another degeneracy is the one between $\alpha$ and $n_s$: if $\alpha$
increases, so does $n_s$. Indeed when $\alpha$ increases, we replace a
part of the contribution of the adiabatic spectrum by a contribution
of topological defects. But, at large scales, $P(k)$ due to
topological defects is more concave than $P(k)$ predicted by the
adiabatic model, which can be taken as $P(k) \propto {k}^{n_s-1}$
\cite{Kolb} with $n_s=0.99 \pm 0.04$ \cite{WMAPbestfit}. Thus, if we
increase the contribution of topological defects, we must increase the
convexity of the adiabatic spectrum to keep a shape compatible with
WMAP results, that is to say that $n_s$ has to increase, which it does. 

If we use the set of parameters corresponding to the maximum
likelihood to compute the power spectrum, we find a spectrum in
excellent agreement with WMAP best fit: the first two peaks coincide
perfectly. Nevertheless, a small discrepancy can be seen for the third
peak. This is consistent with the model of the power spectrum induced
by topological defects which gives a quasi-zero contribution after the
second peak. Therefore, if the contribution of the adiabatic spectrum
decreases, it is not compensated by the added contribution of
topological defects, which makes the third peak decrease. Despite this
feature, the fit is still compatible with WMAP results. It should also
be noticed that this is true for each value of $\alpha$ included in
the 1-$\sigma$ and 2-$\sigma$ contours; small discrepancies begin to
appear in the 3-$\sigma$ area.

During this work, we became aware of the paper \cite{Pogosian}, which
uses a Bayesian analysis in a three dimensional parameter space to
constrain the contribution of cosmic strings with first year WMAP
data. In their work, $\omega_m$, $\omega_b$, $h$ and $\tau$ are chosen
to be WMAP best fit values, and they study the likelihood of their
model involving cosmic strings in a 3-D parameters space, the only
non-topological parameter involved in this space being $n_s$. The
results they find with this method are similar to ours concerning the
best fit, but quite different as far as contours are concerned, and
this can be easily explained. Indeed, in \cite{Pogosian} almost all
cosmological parameters are kept constant, at the exception of $n_s$,
with their values corresponding to WMAP best fit, which the authors
claim is allowed by the fact that the contribution of topological
defects should be relatively small compared to the one of quantum
fluctuations. This is absolutely true for the best fit which gives a
contribution of cosmological defects of order $4\%$, the other
cosmological parameters being very close to those estimated by the
NASA/WMAP Science Team, as we have shown in this Letter. Nevertheless,
this statement is wrong generally speaking. Indeed, we have seen that
letting all parameters free to vary in the parameter space, shows
strong degeneracies between some cosmological parameters (namely $h$,
$\omega_b$, $n_s$ and $\alpha$) so that it is not possible to choose a
parameter independently from the others. Doing so necessarily leads to
wrong estimations of contours, which explain the differences between
our results and those given in \cite{Pogosian}.  

As we were ready to submit this Letter, \cite{wyman}, in which a work
similar to ours is conducted, was posted and gives results consistent
with our work. However, in this paper, the authors use the power
spectrum induced by local cosmic strings, instead of the global
strings we consider. Its shape corresponds to the one we display on
Fig. \ref{fig:powerspectra}, but the peak that is located around $\ell
\sim 150$ on our power spectrum is to be found at $\ell \sim 450$ in
\cite{wyman}. The interesting point is that, even with this
difference, we still get similar results. This suggests that the
results obtained by the method we both use is qualitatively
insensitive to the model of cosmic strings we consider.

\begin{center}
\textbf{IV. \quad CHECKS}
\end{center}

There are two main risks when using a MCMC. First of all, we have to
make sure that the chain converges properly, that is to say that we
have reached a situation in which the likelihood oscillates with a
small amplitude around a mean value which will be the center of the
most likely region for the cosmological parameters to be. Moreover, as
the chain is necessarily finished, we cannot have access to all areas
of the 7-D parameter space. But we have to be sure that we
sufficiently explore this space, in other words, that we do not forget
the exploration of entire areas of the parameter space.

To check these two aspects, we use the approach described in
\cite{Verde} and use the method of \cite{gelman}, which states that
the MCMC gives reliable results if the quantity
\begin{equation}
R=\left(1-\frac{1}{n_e}\right)+\frac{B}{W}\,\left(1+\frac{1}{n_c}\right)
\end{equation}
where $W$ is the within-chain variance and $B$ the between-chain
variance, is smaller than 1.2 when computed on $n_c$ chains each of
which contains $2 n_e$ elements, from which we consider only the last
$n_e$. In this work, we are more conservative and we have chosen to
require $R \sim 1.1$. This result is obtained with 10 chains of 100,000
steps. We also discard the first 5,000 points of each chain, to
eliminate the so-called \textit{burn-in} zone, during which the
stationary distribution might not be reached. The results are
absolutely not sensitive to this last choice, which is an additional
clue showing that we consider enough points in the stationary
distribution.

\begin{center}
\textbf{V. \quad CONCLUSION}
\end{center}

The first year data of WMAP enable to reject at $3\mbox{-}\sigma$
scenarios involving a contribution of more than 29\% of topological
defects, and so, in particular, of cosmic strings. In other words, D-
and F-term inflationary models involving a contribution of
more than 50\% \cite{Dterm} do not match WMAP data, and so, cannot be
acceptable scenarios. But most D- and F-term inflationary models predict
more reasonable contributions of defects. For
example, Urrestilla, Ach\'ucarro \& Davis \cite{Dwithoutcs} present a
superstring-inspired D-term inflation scenario which does not lead to
the formation of cosmic strings. Another example is Rocher \&
Sakellariadou \cite{Rocher} who show that in the context of global
supersymmetry, for F-term inflation, and local supersymmetry, for
D-term inflation, these models can lead to contributions compatible
with our study.

We have also shown in this Letter that we cannot reject models
involving a low contribution (between 2\% and 15\%) of topological
defects, the best fit of WMAP data being given by a 4\% contribution
of topological defects, which gives a result as acceptable as a simple
$\Lambda$CDM model. Concerning this point, this work goes in the same
way as the results obtained by Bouchet \textit{et al.} \cite{Bouchet}
with BOOMERanG data, despite the fact that our best fit gives a much
smaller contribution for cosmic strings thanks to the precision of
WMAP measurements.

Thus, looking at large scales is not sufficient to rule out or confirm
models involving topological defects for the moment. But as WMAP will
soon get more precise measurements of the CMB anisotropies for the
third peak, it will be possible to increase the constraints
on $\alpha$. It should also be worth looking at what topological
defects imply at small scales. Experimental data usable for these
scales will indeed soon be provided by some space or ground based
experiments such as Planck or ACT, so that we will be able to compare
theoretical predictions with observations. 

\begin{acknowledgments}

\begin{center}
\textbf{Acknowledgments}
\end{center}

We thank David Spergel for his guidance through this work and his
support of the project. Olivier Dor\'e is acknowledged for many
useful discussions and suggestions. Hiranya Peiris and Licia Verde
gave precious help in the use and adaptation of CMBwarp. 
The work of AAF is supported by NASA through the Astrophysics Theory
Program and \'Ecole Normale Sup\'erieure de Cachan (France). We
acknowledge the use of the Legacy Archive for Microwave 
Background Data Analysis (LAMBDA). Support for LAMBDA is provided by
the NASA Office of Space Science.

\end{acknowledgments}


\begin{thebibliography}{99}

\bibitem{Linde}
	A. Linde, 
	\textit{Particle Physics and Inflationary Cosmology}, 
	(Harwood Academic, New York,~1990).

\bibitem{Dodelson}
        S. Dodelson,
        \textit{Modern Cosmology}
        (Academic Press, 2003).

\bibitem{WMAPbestfit}
	D. N. Spergel \emph{et al.},
	Astrophys. J. Suppl. {\bf 148}, 175 (2003).


\bibitem{Vilenkin}
	A. Vilenkin and E. P. S. Shellard,
	\textit{Cosmic Strings and Other Topological Defects}
	(Cambridge University Press, 1994).

\bibitem{Srivastava}
        A. M. Srivastava,
        Pramana-J. Phys. {\bf 53}, 1069 (1999).

\bibitem{Kibble}
	T. W. B. Kibble, 
	Phys. Rept. {\bf 67}, 183 (1980).

\bibitem{Kolb}
	E. W. Kolb and M. S. Turner,
	\textit{The Early Universe}
	(Westview Press, 1994).

\bibitem{Dterm}
	R. Kallosh and A. Linde, 
	JCAP {\bf 0310}, 008 (2003);
	R. Jeannerot, 
	Phys. Rev. D {\bf 56}, 6205~(1997). 

\bibitem{Rocher}
        J. Rocher and M. Sakellariadou,
	hep-ph/0406120.

\bibitem{Bouchet}
	F. R. Bouchet, P. Peter, A. Riazuelo, and M. Sakellariadou, 
	Phys. Rev. D. {\bf 65}, 021301 (2001).

\bibitem{seljak}
	U. Seljak, U.-L. Pen, and N. Turok, 
	Phys. Rev. Lett. {\bf 79}, 1615 (1997).

\bibitem{seljaktest}
	N. Turok, U.-L. Pen, and U. Seljak, 
	Phys. Rev. D. {\bf 58}, 023506 (1997).

\bibitem{allen}
	B. Allen \ea, 
	Phys. Rev. Lett. {\bf 79}, 2624 (1997).

\bibitem{cmbwarp}
	R. Jimenez, L. Verde, H. V. Peiris, and A. Kosowsky,
	Phys. Rev. D {\bf 70}, 023005 (2004).

\bibitem{christensen}
	N. Christensen and R. Meyer,
	astro-ph/0006401

\bibitem{Verde}
        L. Verde \textit{et al.},
	Astrophys. J.  Suppl. {\bf 148}, 195 (2003).

\bibitem{Pogosian}
        L. Pogosian, M. Wyman and I. Wasserman,
	JCAP {\bf 09}, 008 (2004).

\bibitem{wyman}
        M. Wyman, L. Pogosian and I. Wasserman,
	astro-ph/0503364.

\bibitem{gelman}
	A. Gelman and D. Rubin,
	Stat. Sci. {\bf 7}, 457 (1992).

\bibitem{Dwithoutcs}
	J. Urrestilla, A. Ach\'ucarro and A. C. Davis,
	Phys. Rev. Lett. {\bf 92}, 251302 (2004).

\end{thebibliography}
\end{document}